\documentclass[preprint,showpacs,onecolumn,nofootinbib]{revtex4}
\pdfoutput=1
\usepackage[colorlinks=true,linkcolor=blue,urlcolor=blue,filecolor=black,citecolor=red,pdfstartview=FitV,pdftitle={},pdfsubject={},pdfkeywords={},pdfpagemode=None,bookmarksopen=true]{hyperref}
\usepackage{graphicx}
\usepackage{amsfonts}
\usepackage{amssymb,ulem}
\usepackage{color}%
\usepackage{dcolumn}
\usepackage{amsmath}
\usepackage{epsfig}
\usepackage{graphics}
\usepackage[active]{srcltx}
\usepackage{epstopdf}
\usepackage{shuffle}
\usepackage[utf8]{inputenc}

\newcommand{\bea}{\begin{eqnarray}}
\newcommand{\eea}{\end{eqnarray}}
\newcommand{\bean}{\begin{eqnarray*}}
\newcommand{\eean}{\end{eqnarray*}}
\newcommand{\nn}{\nonumber \\}

\def\W #1{\widetilde{#1}}

\def\eref#1{(\ref{#1})}

\def\Label#1{\label{#1}%
  \smash{\hbox to0pt{\raise1ex\hbox{\tiny[#1]}\hss}}}

\usepackage{MnSymbol,wasysym}
\usepackage{braket}
\usepackage{color}%
\usepackage{eurosym}
\usepackage{calrsfs}
\usepackage[usenames,dvipsnames,svgnames]{xcolor}

\begin{document}

\baselineskip=0.6 cm
\title{Expansion of tree amplitudes for EM and other theories}
\author{Shi-Qian Hu}
\email{mx120170256@yzu.edu.cn}
\author{Kang Zhou}
\email{zhoukang@yzu.edu.cn}

\affiliation{Center for Gravitation and Cosmology, College of Physical Science and Technology, Yangzhou University, Yangzhou, 225009, China}

\begin{abstract}
\baselineskip=0.6 cm
The expansions of tree-level
amplitudes for one theory into amplitudes for another theory, which have been studied in various recent literatures, exhibit hidden connections between different
theories that are invisible in traditional Lagrangian formulism of quantum field theory.
In this paper, the general expansion of tree EM (Einstein-Maxwell) amplitudes
into KK basis of tree YM (Yang-Mills) amplitudes have been derived by applying the method based on differential operators.
The obtained coefficients are shared by the expansion of tree $\phi^4$ amplitudes into tree BS (bi-adjoint scalar) amplitudes, the expansion of tree sYMS (special Yang-Mills-scalar) amplitudes into tree BS amplitudes, as well the expansion of tree DBI (Dirac-Born-Infeld) amplitudes into tree special extended DBI amplitudes.

\end{abstract}

\maketitle


\section{Introduction}

The modern researches on S-matrix have exhibited remarkable properties of scattering amplitudes
that are not evident up on inspecting traditional Feynman rules. The expansion of tree-level amplitudes of one theory into that of another
theory, which hints the hidden connections between different theories, is a significant example. Such unexpected expansions
was first implied by the well known Kawai-Lewellen-Tye (KLT) relation \cite{Kawai:1985xq}, which represents the tree-level amplitudes of gravity as the double copy of color-ordered tree amplitudes of Yang-Mills (YM) theory,
\bea
\mathrm{A}_{\rm G}=\sum_{\sigma,\W\sigma\in S_{n-3}}\,\mathrm{A}^{L}_{\rm YM}(n-1,n,\sigma,1)\mathrm{S}[\sigma|\W\sigma]
\mathrm{A}^{R}_{\rm YM}(1,\W\sigma,n-1,n)\,,~~~~\label{KK-relation}
\eea
where $S_{n-3}$ denotes permutations on $n-3$ external gluons and ${\mathrm S}[\sigma|\W\sigma]$ stands for the kinematic kernel.
One can arrive the expansion of gravitational amplitudes into YM ones by summing over all $S_{n-3}$ permutations $\sigma$ (or $\W\sigma$) in the relation \eref{KK-relation}. The Cachazo-He-Yuan (CHY) formula proposed in 2013 indicates a much richer web of expansions among a wide range of theories \cite{Cachazo:2013gna,Cachazo:2013hca, Cachazo:2013iea, Cachazo:2014nsa,Cachazo:2014xea}. In the CHY framework, the $n$-point tree-level amplitudes arise as a multi-dimensional contour integral over $n$ auxiliary complex variables, formally written as
\bea
\mathrm{A}_n=\int\,d\mu_n\,{I}^{\rm CHY}\,,
\eea
where the auxiliary variables are fully localized by the scattering equations.
The integrand ${ I}^{\rm CHY}$ depends on the theory under consideration, which can always be factorized as two ingredients
\bea
{ I}^{\rm CHY}={ I}^L\times{ I}^R\,.
\eea
Integrands for Gravity and YM take the formulae ${ I}^L_{\rm G}={\rm Pf}'\Psi(k_i,\epsilon_i,z_i)$, ${ I}^R_{\rm G}={\rm Pf}'\Psi(k_i,\W\epsilon_i,z_i)$, and ${ I}^L_{\rm YM}={\rm Pf}'\Psi(k_i,\epsilon_i,z_i)$, ${ I}^R_{\rm YM}={\rm PT}(\alpha)$ (or
${ I}^L_{\rm YM}={\rm PT}(\alpha)$, ${I}^R_{\rm YM}={\rm Pf}'\Psi(k_i,\W\epsilon_i,z_i)$), respectively. Here $\Psi(k_i,\epsilon_i,z_i)$
is a $2n\times 2n$ anti-symmetric matrix depends on external momenta $k_i$, polarization vectors $\epsilon_i$, as well as auxiliary variables
$z_i$. ${\rm Pf}'\Psi(k_i,\epsilon_i,z_i)$ denotes the reduced pfaffian of the matrix $\Psi(k_i,\epsilon_i,z_i)$. ${\rm PT}(\alpha)$ is the so-called
Parke-Taylor factor writes
\bea
{\rm PT}(\alpha)={1\over (z_{\alpha_1}-z_{\alpha_2})(z_{\alpha_2}-z_{\alpha_3})\cdots(z_{\alpha_n}-z_{\alpha_1})}\,.
\eea
Notice that the polarization tensors of gravitons are expressed as the product of polarization vectors $\epsilon_i^\mu\W\epsilon_i^\nu$.
In the framework, the expansion can be understood as expanding the reduced pfaffian ${\rm Pf}'\Psi(k_i,\W\epsilon_i,z_i)$ into the Parke-Taylor
factors ${\rm PT}(\alpha)$. One can go further since the CHY integrands reflect the double copy structure for a verity of other theories beyond gravity. For instance, integrands
for Einstein-Yang-Mills theory (EYM), Einstein-Maxwell theory (EM), Born-Infeld theory (BI) also carry ${ I}^L={\rm Pf}'\Psi(k_i,\epsilon_i,z_i)$ \cite{Cachazo:2014xea}, thus amplitudes of these theories can also be expanded into YM ones by expanding ${ I}^R$ into Parke-Taylor factors.

Although conceptually simple and straightforward, the evaluation of the explicit coefficients of expansion is rather complicate.
 In order to overcome the inadequacy, various methods have been developed from different angles \cite{Stieberger:2016lng,Schlotterer:2016cxa,Chiodaroli:2017ngp,DelDuca:1999rs,Nandan:2016pya,delaCruz:2016gnm,Fu:2017uzt,Teng:2017tbo,Du:2017kpo,Du:2017gnh,Feng:2019tvb}.
Among these investigations, the recently proposed approach based on differential operators is one of the efficient and systematic way \cite{Feng:2019tvb}. The differential operators introduced by  Cheung, Shen and Wen \cite{Cheung:2017ems} link on-shell tree amplitudes for a series of theories together, and organize them
into an elegant unified web\footnote{A similar web for CHY integrands of
various theories has been provided earlier in \cite{Cachazo:2014xea}, and the relation between two pictures was established in \cite{Zhou:2018wvn,Bollmann:2018edb}.}. In this unified web, tree amplitudes of other theories can be generated by acting proper differential operators
on tree amplitudes of gravity, thus the connections between amplitudes of different theories can be represented by differential equations.
With these relations, expansions of tree amplitudes can be arrived via two paths. One is solving the corresponding
differential equations,
together with considering some physical constraints such as gauge invariance condition. Through the manipulation, the expansion of tree
amplitudes for EYM and gravity into tree pure YM amplitudes\footnote{More precisely, the expansion of amplitudes of EYM and gravity into KK basis.
The KK basis will be explained in the next section.} have been derived efficiently \cite{Feng:2019tvb}. {Another way is getting the expansion of other amplitudes by applying corresponding differential operators on the given expansion of gravitational amplitudes since amplitudes of other theories
arise from amplitudes of gravity via differential operators.} Along this path, the expansion of BI amplitudes into YM ones have been obtained directly,
and the expansion of YM amplitudes into bi-adjoint scalar (BS) ones has also been discussed \cite{Feng:2019tvb}. It is worth to emphasize that the connections between amplitudes of different theories, which are reflected by differential operators previously, are now established by expansions.

The unified web based on differential operators includes much more theories than gravity, EYM, BI and YM, thus it is natural to expect that
the same method can be applied to other theories in the web, and expansions for these theories could also be found. The current short paper devote to apply the method based on differential operators to tree amplitudes of Einstein-Maxwell theory (EM), $\phi^4$ theory, special Yang-Mills-scalar theory (sYMS), as well as Dirac-Born-Infeld theory (DBI). By applying appropriate operators on the expansion of tree gravitational amplitudes, it is simple to obtain the expansion of tree EM amplitudes into KK basis. Then, the expansion of tree $\phi^4$ amplitudes into tree BS amplitudes, the expansion of tree sYMS amplitudes into tree BS amplitudes, and the expansion of tree DBI amplitudes into tree special extended DBI (seDBI) amplitudes, can be readout straightforwardly by acting operators on the expansion of EM amplitudes. Thus the web of the expansions which reflects the deep connections between different theories have been
generalized to include EM, YM, $\phi^4$, sYMS, BS, DBI as well as seDBI.

This paper is organized as follows. In section \ref{review}, we give a brief review of differential operators and the formula of the expansion of gravitational amplitudes into KK basis. These ingredients serve as backgrounds for the work in this paper. With these preparations, we derive the general expression of the expansion of tree EM amplitudes into KK basis in section \ref{general-form}. As by products, the expansion of $\phi^4$ amplitudes into BS ones, the expansion of sYMS amplitudes into BS ones, and the expansion of DBI amplitudes into special extended DBI ones, are also provided in this section. Some explicit examples are given in section \ref{example}. Finally, we end with a summary and discussions in section \ref{conclusion}.

\section{Back grounds}
\label{review}

In this section, we review some already known results, which are crucial for discussions in later sections.
Firstly, we introduce differential operators which are main tools in this paper. Secondly, we review the KK basis
of pure YM amplitudes, and the explicit
formula of the expansion of tree gravitational amplitudes into KK basis.

\subsection{Differential operators}

The differential operators defined by cheung, Shen and Wen, act on variables constructed by Lorentz contractions of
external momenta and polarization vectors,  establish unifying relations for a variety of theories \cite{Cheung:2017ems,Zhou:2018wvn,Bollmann:2018edb}. There are three kinds of basic operators
\begin{itemize}
\item (1) Trace operator:
\bea
{\rm T}^\epsilon_{ij}\equiv {\partial\over\partial(\epsilon_i\cdot\epsilon_j)}\,,
\eea
where $\epsilon_i$ is the polarization vector of $i$th external leg.
\item (2) Insertion operator:
\bea
{\rm T}^\epsilon_{ikj}\equiv {\partial\over\partial(\epsilon_k\cdot k_i)}-{\partial\over\partial(\epsilon_k\cdot k_j)}\,,
\eea
where $k_i$ denotes the momentum of the $i$th external leg.
\item (3) Longitudinal operator:
\bea
{ L}^\epsilon_i\equiv \sum_{j\neq i}\,k_i\cdot k_j{\partial\over\partial(\epsilon_i\cdot k_j)}\,,~~~~~~~~
{ L}^\epsilon_{ij}\equiv -k_i\cdot k_j{\partial\over\partial(\epsilon_i\cdot \epsilon_j)}\,.
\eea
\end{itemize}
Here the up index $\epsilon$ means the operators are defined through polarization vectors $\epsilon_i$.
For gravitons with two copies of polarization vectors, operators can also be defined via another independent copy $\W\epsilon_i$.

Four kinds of combinatory operators can be defined as products of these basic operators
\begin{itemize}
\item (1) For a length-$m$ ordered set $\alpha=\{\alpha_1,\alpha_2,\cdots,\alpha_m\}$ of external particles, the operator ${\rm T}^\epsilon[\alpha]$
is given as
\bea
{\rm T}^\epsilon[\alpha]\equiv {T}^\epsilon_{\alpha_1\alpha_m}\cdot\prod_{i=2}^{m-1}\,{\rm T}^\epsilon_{\alpha_{i-1}\alpha_i\alpha_m}\,,
\eea
which generates the color-ordering $(\alpha_1,\alpha_2,\cdots,\alpha_m)$.
\item (2) For $n$ external particles, the operator ${ L}^\epsilon$
is defined as
\bea
{L}^\epsilon\equiv\prod_i\,{ L}^\epsilon_i,~~~~~~~~\W{ L}^\epsilon\equiv\sum_{\rho\in{\rm pair}}\,\prod_{i,j\in\rho}\,{ L}^\epsilon_{ij}\,.
\eea
Two definitions ${ L}^\epsilon$ and $\W{ L}^\epsilon$ are not equivalent to each other at the algebraic level. However, when acting on appropriate on-shell
physical amplitudes, two combinations
${ L}^\epsilon\cdot{\rm T}^\epsilon_{ab}$ and $\W{ L}^\epsilon\cdot{\rm T}^\epsilon_{ab}$, with subscripts of ${ L}^\epsilon_i$ and ${L}^\epsilon_{ij}$
run through all nodes in $\{1,2,\cdots,n\}/\{a,b\}$, give the same effect and have meaningful physical interpretation.
\item (3) For a length-$2m$ set $I$, the operator ${\rm T}^\epsilon_{X_{2m}}$ is defined by
\bea
{\rm T}^\epsilon_{X_{2m}}\equiv\sum_{\rho\in{\rm pair}}\,\prod_{i_k,j_k\in\rho}\,{\rm T}^\epsilon_{i_kj_k}\,.
\eea
\item (4) The operator ${\rm T}^\epsilon_{{\cal X}_{2m}}$ is defined as
\bea
{\rm T}^\epsilon_{{\cal X}_{2m}}\equiv\sum_{\rho\in{\rm pair}}\,\prod_{i_k,j_k\in\rho}\,\delta_{I_{i_k}I_{j_k}}{\rm T}^\epsilon_{i_kj_k}\,,
\eea
where $\delta_{I_{i_k}I_{j_k}}$ forbids the interaction between particles with different flavors.
\end{itemize}
The explanation for the notation $\sum_{\rho\in{\rm pair}}\,\prod_{i_k,j_k\in\rho}$ is necessary. One can divide the length-$2m$
set $I$ into $m$ pairs, then the set $\rho$ of pairs for this partition can be written as
\bea
\rho=\{(i_1,j_1),(i_2,j_2),\cdots,(i_m,j_m)\}\,,
\eea
with conditions $i_i<i_2<\cdots<i_m$ and $i_t<j_t,\,\forall t$. Under such partitions, $\prod_{i_k,j_k\in\rho}$ stands for the product of ${\rm T}^\epsilon_{i_kj_k}$
for all pairs $(i_k,j_k)$ in $\rho$, and $\sum_{\rho\in{\rm pair}}$ denotes the summation over all possible partitions.

The combinatory operators above translate the tree amplitudes of gravity into tree amplitudes of other theories. Here we only focus on
the generated theories which will be considered later, and list them in Table \ref{tab:unifying}.
\begin{table}[!h]
    \begin{center}
        \begin{tabular}{c|c}
            Amplitude& Operator acts on ${\rm A}^{\epsilon,\W\epsilon}_{{\rm G}}(\{h\}_n)$ \\
            \hline
            ${\rm A}^{\epsilon,\W\epsilon}_{{\rm EM}}(\{p\}_{2m};\{h\}_{n-2m})$ & ${\rm T}^{\W\epsilon}_{X_{2m}}$\\
            ${\rm A}^{\epsilon,\W\epsilon}_{{\rm EMf}}(\{p\}_{2m};\{h\}_{n-2m})$&  ${\rm T}^{\W\epsilon}_{{\cal X}_{2m}}$ \\
            ${\rm A}_{\phi^4}(i'_1,\cdots,i'_n)$ &  ${\rm T}^{\epsilon}[i'_1\cdots i'_n]\cdot {\rm T}^{\W\epsilon}_{X_n}$ \\
            ${\rm A}^{\W\epsilon}_{{\rm sYMS}}(\{s\}_{2m};\{g\}_{n-2m}|i'_1,\cdots,i'_n)$ &  ${\rm T}^{\epsilon}[i'_1\cdots i'_n]\cdot{\rm T}^{\W\epsilon}_{{\cal X}_{2m}}$ \\
            ${\rm A}^{\W\epsilon}_{{\rm DBI}}(\{s\}_{2m};\{p\}_{n-2m})$ &$\Big({ L}^{\epsilon}\cdot{\rm T}^{\epsilon}[ab]\Big)\cdot{\rm T}^{\W\epsilon}_{{\cal X}_{2m}}$ \\
            ${\rm A}^{\epsilon}_{{\rm YM}}(i_1,\cdots,i_n)$ &${\rm T}^{\W\epsilon}[i_1\cdots i_n]$ \\
            ${\rm A}_{{\rm BS}}(i_1,\cdots,i_n|i'_1,\cdots,i'_n)$ &  ${\rm T}^{\epsilon}[i'_1\cdots i'_n]\cdot{\rm T}^{\W\epsilon}[i_1\cdots i_n]$ \\
            ${\rm A}_{{\rm seDBI}}(i_1,\cdots,i_n)$ &  $\Big({ L}^{\epsilon}\cdot{\rm T}^{\epsilon}[ab]\Big)\cdot{\rm T}^{\W\epsilon}[i_1\cdots i_n]$ \\
        \end{tabular}
    \end{center}
    \caption{\label{tab:unifying}Unifying relations}
\end{table}
In this table, EMf stands for EM amplitudes that photons carry flavors, seDBI denotes the special single-trace
extended DBI amplitudes that all external particles are scalars. Sets $\{h\}_a$, $\{p\}_a$, $\{g\}_a$ and $\{s\}_a$
denote gravitons, photons, gluons and scalars respectively, where $a$ is the length of the set.
The notation $(\cdots|i'_1,\cdots,i'_n)$ stands for the additional color-ordering $(i'_1,\cdots,i'_n)$ among all external particles. In the notation
$(\{\alpha\};\{\beta\})$, $\{\alpha\}$ are particles with lower spin, while $\{\beta\}$ are particles with higher spin. The up index of
${\rm A}$ denotes the polarization vectors of external particles by the following rule: $\W\epsilon$ are only carried by particles in $\{\beta\}$,
while $\epsilon$ are carried by all particles.

\subsection{KK basis and expansion of gravitational amplitudes}

When expanding an amplitude into other amplitudes, a necessary requirement is that the basis is complete and linear independent.
Such condition can not be satisfied by all color-ordered tree YM amplitudes with the same external particles, due to the well known
Kleiss-Kuijf (KK) relation \cite{Kleiss:1988ne}
\bea
{\rm A}^{\epsilon}_{\rm YM}(1,\{a\},n,\{b\})=\sum_{\shuffle}\,(-)^{|b|}{\rm A}^{\epsilon}_{\rm YM}(1,\{a\}\shuffle\{b\}^T,n)\,,~~~~\label{KK}
\eea
where $\{a\}$ and $\{b\}$ are two subsets of external gluons, $|b|$ stands for the length of $\{b\}$,
and $\{b\}^T$ denotes the set obtained from $\{b\}$ by reserving the original ordering. The summation is over all
possible shuffles of two ordered subsets, i.e., all permutations of $\{a\}\cup\{b\}$ while preserving the ordering
of $\{a\}$ and $\{b\}$. For example,
\bea
\sum_{\shuffle}\,{\rm A}^\epsilon_{\rm YM}(1,\{3,4\}\shuffle\{5,6\},7)&=&{\rm A}^\epsilon_{\rm YM}(1,3,4,5,6,7)+{\rm A}^\epsilon_{\rm YM}(1,3,5,4,6,7)+{\rm A}^\epsilon_{\rm YM}(1,3,5,6,4,7)\nn
& &+{\rm A}^\epsilon_{\rm YM}(1,5,3,4,6,7)+{\rm A}^\epsilon_{\rm YM}(1,5,3,6,4,7)+{\rm A}^\epsilon_{\rm YM}(1,5,6,3,4,7)\,.
\eea
The KK relation \eref{KK} indicates that an arbitrary color-ordered YM amplitude can be expanded into color-ordered YM amplitudes with two external legs are fixed at two ends. In this sense, $(n-2)!$ amplitudes with two legs fixed
at two ends can be chose as the complete basis (so-called KK basis).
Actually, the KK basis is not independent to each other due to the Bern-Carrasco-Johansson (BCJ) relations among them \cite{Bern:2008qj}.
The BCJ relations suggest the truly independent basis which include $(n-3)!$ amplitudes with three legs are fixed. However, as discussed in \cite{Feng:2019tvb}, the coefficients contain poles when expanding to BCJ basis. If we hope the coefficients are rational functions
of Lorentz invariance kinematic variables and all physical poles are included in the basis, the only candidate is the KK basis. In other words, for rational coefficients, the KK basis is independent.

With the KK basis introduced above, now we discuss the expansion of tree gravitational amplitudes into such basis.
The expansion of gravitational amplitudes into color-ordered YM ones in the framework of ordered splitting is given as \cite{Feng:2019tvb}
\bea
{\rm A}^{\epsilon,\W\epsilon}_{\rm G}(\{h\}_n)=\sum_{\{{\rm Or}_l\}}\,\sum_{\shuffle}\,
\Big(F_0\prod_{l=1}^t\,L_l\Big){\rm A}^\epsilon_{\rm YM}(1,{\rm Or}'_0\shuffle {\rm Or}_1\shuffle\cdots\shuffle{\rm Or}_t,n)\,,
~~~~\label{exp-EM-exp-G2}
\eea
where elements in $\{h\}_n$ are labeled as $\{h\}_n=\{1,2,\cdots,n\}$. Some explanations are in order. The ordered splitting for $n$ elements used here is a little different from that used for EYM amplitudes in \cite{Fu:2017uzt,Teng:2017tbo,Du:2017gnh}, { which can be defined as follows}. First, a reference ordering for elements should be given, for instance $n\prec n-1\prec\cdots\prec 1$. For the current case, $n$
must be fixed at the lowest position. Once the reference ordering
is fixed, the ordered splitting is defined by the ordered set of subsets $\{{\rm Or}_0,{\rm Or}_1,\cdots,{\rm Or}_t\}$ subsequently, satisfying
following conditions:
\begin{itemize}
\item Each subset ${\rm Or}_l\subset\{h\}_n$ is ordered.
\item ${\rm Or}_0\cup{\rm Or}_1\cup\cdots\cup{\rm Or}_t=\{h\}_n$.
\item ${\rm Or}_0=\{1,\cdots,n\}$, i.e., $1$ and $n$ belong to ${\rm Or}_0$ and are fixed at the first and last positions respectively. Notice that two gravitons $1$ and $n$ can be chosen  arbitrary in the set $\{h\}_n$.
\item Denoting $h_l$ as the last element in ${\rm Or}_l$, then $h_0\prec h_1\prec\cdots\prec h_t$ according to the reference ordering (it fixes the ordering of subsets ${\rm Or}_l$
    in the set $\{{\rm Or}_0,{\rm Or}_1,\cdots,{\rm Or}_t\}$).
\item In each subset ${\rm Or}_l$, $h_l$ must satisfy $h_l\prec \alpha$ for arbitrary $\alpha\in {\rm Or}_l,\,\alpha\neq h_l$, according to the reference ordering. On the other hand, there is no requirement for ordering of all other elements.
\end{itemize}
The subset
${\rm Or}'_0$ occurs in \eref{exp-EM-exp-G2} is defined by ${\rm Or}'_0\equiv {\rm Or}_0/\{1,n\}$. With the ordered splitting is given,
$L_l$ for a length-$r$ subset ${\rm Or}_l=\{\gamma_1,\gamma_2,\cdots \gamma_r\}$
is defined as
\bea
L_l=\W\epsilon_{\gamma_r}\cdot f_{\gamma_{r-1}}\cdots f_{\gamma_2}\cdot f_{\gamma_1}\cdot Z_{\gamma_1}\,,
\eea
while $F_0$ for ${\rm Or}'_0=\{\gamma_1,\gamma_2,\cdots \gamma_r\}$ is given by
\bea
F_0=\W\epsilon_{h_1}\cdot f_{\gamma_1}\cdot f_{\gamma_2}\cdots f_{\gamma_{r}}\cdot\W\epsilon_{h_n}\,.
\eea
Here $f_{\gamma_i}^{\mu\nu}$ stands for the strength tensor $f_{\gamma_i}^{\mu\nu}\equiv k_{\gamma_i}^\mu \W\epsilon_{\gamma_i}^\nu-\W\epsilon_{\gamma_i}^\mu k_{\gamma_i}^\nu$. The combinatory momentum $Z_i^\mu$ is the sum of momenta of external legs satisfy two conditions: (1) legs at the LHS of the leg $i$
in the color-ordered YM amplitude, (2) legs at the LHS of $i$ in the labeled chain defined by the ordered splitting. The labeled chain
used here for a given ordered splitting is the ordered set $\{1,{\rm Or}'_0,{\rm Or}_1,\cdots,{\rm Or}_t,n\}$.
The summation over shuffles $\sum_{\shuffle}$ was defined previously when introducing KK basis.
The summation $\sum_{\{{\rm Or}_l\}}$ means summing over all possible ordered splittings.

In the formula \eref{exp-EM-exp-G2}, the gravitational amplitude is expanded into color-ordered YM amplitudes with two legs are fixed at two ends,
thus the coefficients of expansion into KK basis can be readout conveniently. The formula \eref{exp-EM-exp-G2} serves as the starting point for computations
in the next section.

Before ending this subsection, we emphasize that in the expansion \eref{exp-EM-exp-G2}, the basis only carry polarization vectors $\epsilon$,
and all polarization vectors $\W\epsilon$ are included in the coefficients.

\section{Derivation of expansions}
\label{general-form}

In this section, we derive the general formulae of expansions of tree amplitudes.
We first apply operators ${\rm T}^{\W\epsilon}_{X_{2m}}$ and ${\rm T}^{\W\epsilon}_{{\cal X}_{2m}}$ on two sides of \eref{exp-EM-exp-G2}
simultaneously to obtain the ordered splitting formula of the expansion of tree EM amplitudes into YM ones.
Then, we propose the rule of getting the coefficients of KK basis.
Finally, we explain that coefficients of the expansion of gravitational amplitudes into KK basis are shared by the expansion of $\phi^4$ amplitudes into BS ones, the expansion of sYMS amplitudes into BS ones, as well as the expansion of DBI amplitudes into special extended DBI ones.

\subsection{Expansion of EM amplitudes into YM ones: ordered splitting formula}

We start with tree amplitudes of general Einstein-Maxwell theory that photons do not carry any flavor.
To generate such amplitudes, one can act the operator ${\rm T}^{\W\epsilon}_{X_{2m}}$ on gravitational amplitudes, which can be seen in Table \ref{tab:unifying}.
Now we are going to apply this operator on two sides of \eref{exp-EM-exp-G2}.
On the LHS, the operator ${\rm T}^{\W\epsilon}_{X_{2m}}$ transmute the gravitational amplitude into the EM one.
On the RHS, since basis depend on polarization vectors $\epsilon_i$ and all $\W\epsilon_i$ are included in coefficients,
the operator ${\rm T}^{\W\epsilon}_{X_{2m}}$ only modifies coefficients. Thus, the action of operator
provides the expansion of EM amplitudes into pure YM amplitudes, with the coefficients determined by acting the operator on
coefficients in the expansion of gravitational amplitudes.

As a warm up, we first restrict ourselves on the special case that all external particles of the EM amplitude
are photons, i.e., $2m=n$. Under this condition, subscripts $i$ and $j$ of ${\rm T}^{\W\epsilon}_{ij}$ in ${\rm T}^{\W\epsilon}_{X_{2m}}$ run over all external particles. To study the effect of ${\rm T}^{\W\epsilon}_{X_{2m}}$, let us consider the operator $\prod_{i_k,j_k\in\rho}\,{\rm T}^{\W\epsilon}_{i_kj_k}$ under a given partition $\rho$. If a term on the RHS of \eref{exp-EM-exp-G2} contains all $(\W\epsilon_{i_k}\cdot\W\epsilon_{j_k})$ with $i_k,j_k\in\rho$, the operator $\prod_{i_k,j_k\in\rho}\,{\rm T}^{\W\epsilon}_{i_kj_k}$ turns all these $(\W\epsilon_{i_k}\cdot\W\epsilon_{j_k})$ into $1$. Otherwise, the term will be annihilated by $\prod_{i_k,j_k\in\rho}\,{\rm T}^{\W\epsilon}_{i_kj_k}$.
Thus, in each term which can survive under the action of ${\rm T}^{\W\epsilon}_{X_{2m}}$, every polarization vector must contract with another one.
This requirement indicates that not all ordered splittings are allowed, each subsets in the ordered splitting must take even length
according to the observation that each $(\W\epsilon_{i_k}\cdot\W\epsilon_{j_k})$ only occurs in $L_l$ or $F_0$.
Acting ${\rm T}^{\W\epsilon}_{X_{2m}}$ on terms correspond to these selected splittings, one can get the non-vanishing contributions.
For proper ordered splittings with even length, $L_l$ and $F_0$ contain
\bea
(-)^{|{\rm Or}_l|\over 2}(\W\epsilon_{\gamma_r}\cdot\W\epsilon_{\gamma_{r-1}})(k_{\gamma_{r-1}}\cdot k_{\gamma_{r-2}})\cdots (\W\epsilon_{\gamma_4}\cdot\W\epsilon_{\gamma_3})(k_{\gamma_3}\cdot k_{\gamma_2})(\W\epsilon_{\gamma_2}\cdot\W\epsilon_{\gamma_1})(k_{\gamma_1}\cdot Z_{\gamma_1})\,,
\eea
and
\bea
(-)^{{|{\rm Or}_0|\over 2}}(\W\epsilon_{1}\cdot\W\epsilon_{\gamma_1})(k_{\gamma_1}\cdot k_{\gamma_2})(\W\epsilon_{\gamma_2}\cdot\W\epsilon_{\gamma_3})(k_{\gamma_3}\cdot k_{\gamma_4})\cdots (\W\epsilon_{\gamma_{r-2}}\cdot\W\epsilon_{\gamma_{r-1}})(k_{\gamma_{r-1}}\cdot k_{\gamma_{r}})(\W\epsilon_{\gamma_r}\cdot\W\epsilon_n)\,,
\eea
respectively.
After turning all $(\W\epsilon_{i_k}\cdot\W\epsilon_{j_k})$ into $1$, we get the expansion of the tree EM amplitude as
\bea
{\rm A}^{\epsilon,\W\epsilon}_{\rm EM}(\{p\}_n;\emptyset)=\sum_{|{\rm Or}_l|{\rm even}}\,\sum_{\shuffle}\,
\Big(E_0\prod_{l=1}^t\,N_l\Big){\rm A}^\epsilon_{\rm YM}(1,{\rm Or}'_0\shuffle {\rm Or}_1\shuffle\cdots\shuffle{\rm Or}_t,n)\,,
~~~~\label{exp-EM1}
\eea
where
\bea
E_0=(-)^{{|{\rm Or}_0|\over 2}}(k_{\gamma_1}\cdot k_{\gamma_2})(k_{\gamma_3}\cdot k_{\gamma_4})\cdots (k_{\gamma_{r-1}}\cdot k_{\gamma_{r}})\,,
\eea
and
\bea
N_l=(-)^{|{\rm Or}_l|\over 2}(k_{\gamma_{r-1}}\cdot k_{\gamma_{r-2}})\cdots (k_{\gamma_3}\cdot k_{\gamma_2})(k_{\gamma_1}\cdot Z_{\gamma_1})\,.
\eea
If the set $\{{\rm Or}'_0\}$ is empty, we have $E_0=1$.

Then we turn to the general case that the $n$-point tree EM amplitude contains $2m$
photons and $(n-2m)$ gravitons as external particles. Since two special gravitons $1$ and $n$ in \eref{exp-EM-exp-G2} can be chosen arbitrary, we assume that both
$1$ and $n$ are turned into photons. Let us consider a given partition $\rho=\{(i_1,j_1),(i_2,j_2),\cdots,(i_{m},j_{m})\}$
with $i_1<i_2<\cdots <i_{m}$ and $i_k<j_k$, $\forall k$ for external photons.
Under the action of corresponding $\prod_{(i_k,j_k)\in \rho}\,{\rm T}^{\W\epsilon}_{i_kj_k}$
in the operator ${\rm T}^{\W\epsilon}_{X_{2m}}$, all non-vanishing terms on the RHS of \eref{exp-EM-exp-G2} must contain all $(\W\epsilon_{i_k}\cdot\W\epsilon_{j_k})$
with $(i_k,j_k)\in\rho$. It indicates that each pair in $\rho$ should appear in one subset of the ordered splitting
at nearby positions, according to the definition of $L_l$ and $F_0$. More explicitly,
if ${\rm Or}_l$ includes $i_k$, it must take the form $\{\cdots,j_k,i_k,\cdots\}$ or $\{\cdots,i_k,j_k,\cdots\}$. If ${\rm Or}_l$
does not contain $j_k$, or contains $j_k$ but $i_k$ and $j_k$ are not nearby, the corresponding term on the RHS of \eref{exp-EM-exp-G2}
is annihilated by $\prod_{(i_k,j_k)\in \rho}\,{\rm T}^{\W\epsilon}_{i_kj_k}$.
We use the notation ${\rm Or}_l^p$ to denote subsets under these proper ordered splittings.
With the ordered splittings are determined, now we consider the effect of acting the operator on corresponding coefficients.
For ${\rm Or}_0^p$, since $F_0$ includes $\W\epsilon_1^\mu$ and $\W\epsilon_n^\mu$ at the first and last positions, the action of
$\prod_{(i_k,j_k)\in \rho}\,{\rm T}^{\W\epsilon}_{i_kj_k}$ turns the vectors $(\W\epsilon_{1}\cdot f_{a})^\mu$
and $(f_b\cdot\W\epsilon_{n})^\mu$ into $-k_a^\mu$ and $k_b^\mu$, respectively. For other
pairs in ${\rm Or}_0^p$, the tensors $(f_{i_k}\cdot f_{j_k})^{\mu\nu}$ are turned into $-k_{i_k}^\mu k_{j_k}^\nu$.
All $(-)$ signs give rise to $(-)^{|p_0|-1}$, where $|p_0|$ is the number of photon-pairs in ${\rm Or}_0^p$.
Thus for the subset ${\rm Or}_0^p$ we get $(-)^{|p_0|-1}G_0$, where $G_0$ can be obtained from $F_0$
by the replacement
\bea
(\W\epsilon_{1}\cdot f_{a})^\mu\to k_a^\mu,~~~~(f_b\cdot\W\epsilon_{n})^\mu\to k_b^\mu,
~~~~(f_{i_k}\cdot f_{j_k})^{\mu\nu}\to k_{i_k}^\mu k_{j_k}^\nu\,.
\eea
If ${\rm Or}_0^p=\emptyset$, we have $(\W\epsilon_1\cdot\W\epsilon_n)\to1$.
For other ${\rm Or}_l^p$, if $i_k$ in one pair $(i_k,j_k)$ is at the last position
of the subset, i.e., appears as $\{\cdots,j_k,i_k\}$, then the vector $(\W\epsilon_{i_k}\cdot f_{j_k})^\mu$
will be turned into $-k_{j_k}^\mu$. For other cases, the tensors $(f_{i_k}\cdot f_{j_k})^{\mu\nu}$
are turned into $-k_{i_k}^\mu k_{j_k}^\nu$. Thus, one can obtain $(-)^{|p_l|}H_l$, where $H_l$
can be obtained from $L_l$ via the
replacement
\bea
(\W\epsilon_{i_k}\cdot f_{j_k})\to k_{j_k}^\mu,~~~~(f_{i_k}\cdot f_{j_k})^{\mu\nu}\to k_{i_k}^\mu k_{j_k}^\nu\,.
\eea
Collecting these results together, we find the contribution of an individual splitting
can be expressed as
\bea
\sum_{\shuffle}\,(-)^{m-1}\,
\Big(G_0\prod_{l=1}^t\,H_l\Big){\rm A}^\epsilon_{\rm YM}(1,{\rm Or}'^p_0\shuffle {\rm Or}^p_1\shuffle\cdots\shuffle{\rm Or}^p_t,n)\,,
~~~~\label{exp-EM2}
\eea
thus the full expansion is given by
\bea
\begin{split}
{\rm A}^{\epsilon,\W\epsilon}_{\rm EM}(\{p\}_{2m};\{h\}_{n-2m})~~~~~~~~~~~~~~~~~~~~~~~~~~~~~~~~~~~~~~~~~~~~~~~~~~~~~~~~~~~~~~~~~~~~\\
=\sum_\rho\,\sum_{\{{\rm Or}_l^p\}^\rho}\,\sum_{\shuffle}\,(-)^{m-1}\,
\Big(G_0\prod_{l=1}^t\,H_l\Big){\rm A}^\epsilon_{\rm YM}(1,{\rm Or}'^p_0\shuffle {\rm Or}^p_1\shuffle\cdots\shuffle{\rm Or}^p_t,n)\,,
~~~~\label{exp-EM3}
\end{split}
\eea
where the summation $\sum_{\{{\rm Or}_l^p\}^\rho}$ is over all ordered splittings correspond to a special partition $\rho$ of photons,
and $\sum_\rho$ is over all partitions due to the definition of the operator ${\rm T}^{\W\epsilon}_{X_{2m}}$. When all external particles are photons, it can be verified straightforwardly
the general formula \eref{exp-EM3} is reduced to the special one \eref{exp-EM1}.

At the end of this subsection, we discuss the expansion of EM amplitudes that photons carry flavors.
For this case, the operator ${\rm T}^{\W\epsilon}_{X_{2m}}$ is replaced by ${\rm T}^{\W\epsilon}_{{\cal X}_{2m}}$,
and a contraction $(\W\epsilon_{i_k}\cdot\W\epsilon_{j_k})$
is permitted only when two photons carry the same flavor.
The constraints from $\delta_{I_{i_k}I_{j_k}}$ lead to the conclusion only partitions satisfy $\delta_{I_{i_k}I_{j_k}}=1$
for all $i_k,j_k\in\rho$ provide non-vanishing contributions. Thus, the expansion of these amplitudes in the ordered splitting
formula can be obtained from the formula \eref{exp-EM3} by restricting the summation $\sum_\rho$ on proper partitions.

\subsection{Expansion of EM amplitudes into KK basis}
\label{coe-KK}

The formula \eref{exp-EM3} provides the expansion of tree EM amplitudes into pure YM amplitudes in the framework of ordered splitting.
To obtain the expansion into KK basis, one can extract coefficients of KK basis with desired color-ordering from the obtained expansion in the ordered splitting formula. An alternative way is reconstructing corresponding ordered splittings from the desired color-ordering in KK basis
directly, without requiring the given expansion in the ordered splitting formula, as will be discussed in this subsection.
The procedure for the expansion of EYM amplitudes have been provided in \cite{Fu:2017uzt,Teng:2017tbo,Du:2017gnh}.  The manipulation is similar but a little different for EM amplitudes. Now we propose the algorithm.

Assuming the color-ordering in KK basis is
$(1,2,3,\cdots,n-1,n)$\footnote{Since the EM amplitude do not carry any color-ordering, this choice will not lose any generality.}, and the reference ordering is chosen to be $n\prec i_1\prec i_2\prec\cdots\prec i_{n-1}$. Subsequently, for a given partition $\rho=\{(i_1,j_1),\cdots,(i_m,j_m)\}$,
one can determine the corresponding ordered splittings as follows,
\begin{itemize}
\item First step: List all possible ordered subsets ${\rm Or}_0^p=\{1,\gamma_1,\gamma_2,\cdots,\gamma_r,n\}$, respecting the color-ordering in KK basis, i.e., $\gamma_1<\gamma_2<\cdots\gamma_r$. In addition, if a photon is included in ${\rm Or}_0^p$, its partner in $\rho$
    should also be included in ${\rm Or}_0^p$, and positions of two photons in ${\rm Or}_0^p$ are nearby.
\item Second step: For each ${\rm Or}_0^p$, remove its elements in $\{1,2,\cdots,n\}$. Then, for remaining elements in $\{1,2,\cdots,n\}/{\rm Or}_0^p$, we select the lowest element $h_1$ in the reference ordering and construct all possible ordered subsets ${\rm Or}_1^p=\{\gamma'_1,\gamma'_2,\cdots,\gamma'_{r'},h_1\}$, satisfying $h_1\prec\gamma'_i,\,\forall\gamma'_i\in\{\gamma'_1,\gamma'_2,\cdots,\gamma'_{r'}\}$, and regarding the color-ordering in KK basis, i.e., $\gamma'_1<\gamma'_2<\cdots<\gamma'_{r'}<h_1$. The subset ${\rm Or}_1^p$ also satisfy the condition that each photon contained
    in ${\rm Or}_1^p$ occurs nearby its partner in $\rho$.
\item Repeat the second step until the complete ordered splitting is achieved.
\end{itemize}
All proper ordered splittings for all partitions can be found via the manipulation mentioned above. After generating ordered splittings, the coefficient
for the particular YM amplitude ${\rm A}^\epsilon_{\rm YM}(1,2,\cdots,n)$ can be obtained by summing factors $G_0\prod_{l=1}^t\,H_l$
over all correct ordered splittings and all partitions. Some simple examples will be presented in the next section to illustrate the algorithm more explicitly. 

Until now, the expansion of EM amplitudes into KK basis can be formally represented as
\bea
{\rm A}^{\epsilon,\W\epsilon}_{\rm EM}(\{p\}_{2m};\{h\}_{n-2m})=\sum_{\sigma\in S_{n-2}}\,C^{\W\epsilon}(\sigma,m,\rho){\rm A}^\epsilon_{\rm YM}
(1,\sigma_2,\sigma_3,\cdots,\sigma_{n-1},n)\,,~~~~\label{ex-to-KK}
\eea
where $\sigma$ stands for permutations among $(n-2)$ elements. The coefficients $C^{\W\epsilon}(\sigma,m,\rho)$ depend on polarization vectors
$\W\epsilon$, permutation $\sigma$, number of the photon-pairs $m$, as well as allowed partitions $\rho$. Of course, it also depend on external momenta although the dependence is implicit in the formula \eref{ex-to-KK}. Since the dependence on possible partitions, the formula \eref{ex-to-KK} is correct for EM amplitudes whether or not photons carry flavor.
\subsection{Expansions of $\phi^4$, sYMS and DBI amplitudes}

In this subsection, we will identify that the coefficients $C^{\W\epsilon}(\sigma,m,\rho)$ in the expression \eref{ex-to-KK}
are also the coefficients in the expansion of $\phi^4$ amplitudes into BS amplitudes, the expansion of sYMS amplitudes into BS amplitudes, as well as the expansion of DBI amplitudes into special extended DBI amplitudes.

Let us come to $\phi^4$ theory whose amplitudes can be generated by acting the operator ${\rm T}^{\epsilon}[i'_1\cdots i'_n]$
on EM amplitudes ${\rm A}^{\epsilon,\W\epsilon}_{\rm EM}(\{p\}_{n};\emptyset)$ that all external particles are photons without flavor, due to relations
\bea
{\rm A}^{\epsilon,\W\epsilon}_{\rm EM}(\{p\}_{n};\emptyset)={\rm T}^{\W\epsilon}_{X_n}{\rm A}^{\epsilon,\W\epsilon}_{\rm G}(\{h\}_{n}),~~~~{\rm A}_{\phi^4}(i'_1,\cdots,i'_n)={\rm T}^\epsilon[i'_1\cdots i'_n]{\rm T}^{\W\epsilon}_{X_n}{\rm A}^{\epsilon,\W\epsilon}_{\rm G}(\{h\}_{n})
\eea
in Table \ref{tab:unifying}. If one set the LHS of \eref{ex-to-KK} to be ${\rm A}^{\epsilon,\W\epsilon}_{\rm EM}(\{p\}_{n};\emptyset)$, and act the operator ${\rm T}^\epsilon[i'_1\cdots i'_n]$ on two sides of \eref{ex-to-KK}
simultaneously, the LHS gives the $\phi^4$ amplitude ${\rm A}_{\phi^4}(i'_1,\cdots,i'_n)$. For the RHS, since the operator ${\rm T}^\epsilon[i'_1\cdots i'_n]$ is defined via polarization vectors $\epsilon_i$, it only affect on KK basis ${\rm A}^\epsilon_{\rm YM}
(1,\sigma_2,\sigma_3,\cdots,\sigma_{n-1},n)$ and transmute them into BS amplitudes ${\rm A}_{\rm BS}
(1,\sigma_2,\sigma_3,\cdots,\sigma_{n-1},n;i'_1,\cdots,i'_n)$, as shown in Table \ref{tab:unifying}.
Then, we get the expansion of $\phi^4$ amplitudes into BS ones as
\bea
{\rm A}_{\phi^4}(i'_1,\cdots,i'_n)=\sum_{\sigma\in S_{n-2}}\,C(\sigma,m,\rho){\rm A}_{\rm BS}
(1,\sigma_2,\sigma_3,\cdots,\sigma_{n-1},n|i'_1,\cdots,i'_n)\,.~~~~\label{ex-phi4}
\eea
Coefficients $C(\sigma,m,\rho)$ are products of Lorentz contractions of external momenta, which can be understood
as sewing $4$-point vertexes of $\phi^4$ theory into $3$-point vertexes of BS theory by eliminating propagators.

In the above discussion, if we act the operator ${\rm T}^{\epsilon}[i'_1\cdots i'_n]$ on EMf amplitudes
${\rm A}^{\epsilon,\W\epsilon}_{\rm EMf}(\{p\}_{2m};\{h\}_{n-2m})$ that photons carry flavors, the generated amplitudes are
amplitudes of special YM theory which describe the low energy effective action of coincident D-brans.
Thus we also have
\bea
{\rm A}^{\W\epsilon}_{\rm sYMS}(\{s\}_{2m};\{g\}_{n-2m}|i'_1,\cdots,i'_n)=\sum_{\sigma\in S_{n-2}}\,C^{\W\epsilon}(\sigma,m,\rho){\rm A}_{\rm BS}
(1,\sigma_2,\sigma_3,\cdots,\sigma_{n-1},n|i'_1,\cdots,i'_n)\,.~~~~\label{ex-sYMS}
\eea
Notice that the allowed partitions $\rho$ for \eref{ex-phi4} and \eref{ex-sYMS} are different since constraints from $\delta_{I_{i_k}I_{j_k}}$
for the second one. 

Similarly, by acting the operator ${L}^\epsilon\cdot{\rm T}^\epsilon[ab]$ on two sides of \eref{ex-to-KK},
one get the expansion of DBI amplitudes into the seDBI ones as
\bea
{\rm A}^{\W\epsilon}_{\rm DBI}(\{s_{2m};\{p\}_{n-2m}\})=\sum_{\sigma\in S_{n-2}}\,C^{\W\epsilon}(\sigma,m,\rho){\rm A}_{\rm seDBI}
(1,\sigma_2,\sigma_3,\cdots,\sigma_{n-1},n)\,.~~~~\label{ex-DBI}
\eea

If formulae \eref{ex-phi4}, \eref{ex-sYMS} and \eref{ex-DBI} are correct expansions, basis used in them must be complete and independent. Now we explain that there are KK-like relations
among color-ordered BS and seDBI amplitudes, thus this condition is satisfied.
As pointed out in \cite{Feng:2019tvb}, the KK relation can be derived by using differential operators. Indeed, one can regard the KK relation as the inference of the algebraical property of the differential operator ${\rm T}^{\W\epsilon}[\alpha]$.
To see this, we rewrite the operator ${\rm T}^{\W\epsilon}[\alpha]$ for length-$n$ set $[\alpha]$ as
\bea
{\rm T}^\epsilon[\alpha]&\equiv & {T}^\epsilon_{\alpha_1\alpha_n}\cdot\prod_{i=2}^{n-1}\,{\rm T}^\epsilon_{\alpha_{i-1}\alpha_i\alpha_n}\nn
&=&{T}^\epsilon_{\alpha_1\alpha_n}\cdot\Big(\prod_{i=2}^{k}\,{\rm T}^\epsilon_{\alpha_{i-1}\alpha_i\alpha_n}\Big)\cdot\Big(\prod_{j=k+1}^{n-1}\,{\rm T}^\epsilon_{\alpha_{j-1}\alpha_j\alpha_n}\Big)\nn
&=&{T}^\epsilon_{\alpha_1\alpha_n}\cdot\Big(\prod_{i=2}^{k}\,{\rm T}^\epsilon_{\alpha_{i-1}\alpha_i\alpha_n}\Big)\cdot\Big((-)^{n-k-1}\prod_{j=k+1}^{n-1}\,{\rm T}^\epsilon_{\alpha_n\alpha_j\alpha_{j-1}}\Big)\,.
~~~~\label{pro-trace}
\eea
The operator ${T}^\epsilon_{\alpha_1\alpha_n}\cdot\Big(\prod_{i=2}^{k}\,{\rm T}^\epsilon_{\alpha_{i-1}\alpha_i\alpha_n}\Big)$
generates the color-ordering $(\alpha_1,\alpha_2,\cdots,\alpha_k,\alpha_n)$ which is equivalent to $(\alpha_n,\alpha_1,\alpha_2,\cdots,\alpha_k)$
due to the cyclic symmetry, and the operator $\Big((-)^{n-k-1}\prod_{j=k+1}^{n-1}\,{\rm T}^\epsilon_{\alpha_n\alpha_j\alpha_{j-1}}\Big)$
can be interpreted as inserting $\{\alpha_{n-1},\alpha_{n-2},\cdots,\alpha_{k+1}\}$ between $\alpha_n$ and $\alpha_k$ in $(\alpha_n,\alpha_1,\alpha_2,\cdots,\alpha_k)$ \cite{Cheung:2017ems,Zhou:2018wvn,Bollmann:2018edb}. Setting $\alpha_n=1$, $\alpha_k=n$,
$\{a\}=\{\alpha_1,\cdots,\alpha_{k-1}\}$, $\{b\}=\{\alpha_{k+1},\cdots,\alpha_{n-1}\}$, and applying this operator on the $n$-point gravitational
amplitude ${\rm A}^{\epsilon,\W\epsilon}_{\rm G}(\{h\}_n)$, one get the KK relation \eref{KK} immediately. Since
the algebraical relation \eref{pro-trace} is general, it is not surprising that similar relations exist among color-ordered amplitudes
of other theories beyond YM.
Replacing ${\rm A}^{\epsilon,\W\epsilon}_{\rm G}(\{h\}_n)$ in the above derivation by ${\rm T}^\epsilon[i'_1\cdots i'_n]{\rm A}^{\epsilon,\W\epsilon}_{\rm G}(\{h\}_n)$ and ${L}^\epsilon\cdot{\rm T}^\epsilon[ab]{\rm A}^{\epsilon,\W\epsilon}_{\rm G}(\{h\}_n)$,
the KK-like relations for BS amplitudes and seDBI amplitudes can be obtained as
\bea
{\rm A}_{\rm BS}(1,\{a\},n,\{b\}|i'_1,\cdots,i'_n)=\sum_{\shuffle}\,(-)^{|b|}{\rm A}_{\rm BS}(1,\{a\}\shuffle\{b\}^T,n|i'_1,\cdots,i'_n)\,,
\eea
and
\bea
{\rm A}_{\rm seDBI}(1,\{a\},n,\{b\})=\sum_{\shuffle}\,(-)^{|b|}{\rm A}_{\rm seDBI}(1,\{a\}\shuffle\{b\}^T,n)\,.
\eea
Thus, BS amplitudes and seDBI amplitudes with the color-ordering $(1,\{a\}\shuffle\{b\}^T,n)$ share the completeness and independence
of KK basis, therefore can be chosen as proper basis for rational coefficients. Consequently, one can conclude that
formulae \eref{ex-phi4}, \eref{ex-sYMS} and \eref{ex-DBI} are correct expansions for $\phi^4$, sYMS and DBI amplitudes, respectively.

\section{Examples}
\label{example}

Next, we provide some examples to illustrate the expansions obtained in the previous section.
Since the expansion of $\phi^4$ amplitudes into BS amplitudes, the expansion of sYMS amplitudes into BS amplitudes, and the expansion of DBI amplitudes into seDBI amplitudes share the same coefficients with the expansion of EM amplitudes into KK basis, we only consider EM amplitudes in this section.

\subsection{$4$-point EM amplitude ${\rm A}^{\epsilon,\W\epsilon}_{\rm EM}(\{1,2,3,4\};\emptyset)$}

The first example is the $4$-point EM amplitude ${\rm A}^{\epsilon,\W\epsilon}_{\rm EM}(\{1,2,3,4\};\emptyset)$
with all external particles are photons in the set $\{1,2,3,4\}$.
The reference ordering is chosen as $4\prec2\prec 3\prec1$. We first consider the case photons do not carry flavor.
Our result \eref{exp-EM1} requires the length of each subset in the ordered splitting is even,
and ${\rm Or}_0=\{1,\cdots,4\}$, thus only three splittings
$\{\{1,4\},\{3,2\}\}$, $\{\{1,3,2,4\}\}$ and $\{\{1,2,3,4\}\}$ are proper. After evaluating $E_0$ and $N_l$, we get the expansion in the ordered splitting formula
\bea
\begin{split}
{\rm A}^{\epsilon,\W\epsilon}_{\rm EM}(\{1,2,3,4\};\emptyset)~~~~~~~~~~~~~~~~~~~~~~~~~~~~~~~~~~~~~~~~~~~~~~~~~~~~~~~~~~~~~~~~~~~~~~~~~~~~~\\
=-(k_3\cdot Z_3){\rm A}^\epsilon_{\rm YM}
(1,3,2,4)-(k_2\cdot k_3){\rm A}^\epsilon_{\rm YM}
(1,3,2,4)-(k_2\cdot k_3){\rm A}^\epsilon_{\rm YM}
(1,2,3,4)\,.~~~~\label{result1}
\end{split}
\eea

The formula of expansion depends on the choice of KK basis, i.e., the choice of two special legs which are fixed at two ends in the color-ordering, and the choice of reference ordering.
For instance, if we chose the basis ${\rm A}^\epsilon_{\rm YM}
(1,\sigma_3,\sigma_4,2)$, and the reference ordering $2\prec3\prec 4\prec1$, similar manipulation yields
\bea
\begin{split}
{\rm A}^{\epsilon,\W\epsilon}_{\rm EM}(\{1,2,3,4\};\emptyset)~~~~~~~~~~~~~~~~~~~~~~~~~~~~~~~~~~~~~~~~~~~~~~~~~~~~~~~~~~~~~~~~~~~~~~~~~~~~~\\
=-(k_4\cdot Z_4){\rm A}^\epsilon_{\rm YM}
(1,4,3,2)-(k_3\cdot k_4){\rm A}^\epsilon_{\rm YM}
(1,4,3,2)-(k_3\cdot k_4){\rm A}^\epsilon_{\rm YM}
(1,3,4,2)\,.~~~~\label{result2}
\end{split}
\eea
As a verification of self-consistency, we need to prove two expressions \eref{result1} and \eref{result2} are equivalent.
We first use the observation that $Z_3=Z_4=k_1$,
together with the momentum conservation law and on shell condition $k_i^2=0$, to turn
\eref{result1} into
\bea
E_1=(k_3\cdot k_4){\rm A}^\epsilon_{\rm YM}
(1,3,2,4)-(k_2\cdot k_3){\rm A}^\epsilon_{\rm YM}
(1,2,3,4)\,,
\eea
and \eref{result2} into
\bea
E_2=(k_4\cdot k_2){\rm A}^\epsilon_{\rm YM}
(1,4,3,2)-(k_3\cdot k_4){\rm A}^\epsilon_{\rm YM}
(1,3,4,2)\,.
\eea
Using the ordered reserved identity
\bea
{\rm A}^\epsilon_{\rm YM}(i_1,i_2,\cdots,i_n)=(-)^n{\rm A}^\epsilon_{\rm YM}(i_n,i_{n-1},\cdots,i_1)\,,
\eea
together with the cyclic symmetry of color-ordering, we have
\bea
{\rm A}^\epsilon_{\rm YM}(1,4,3,2)={\rm A}^\epsilon_{\rm YM}
(2,3,4,1)={\rm A}^\epsilon_{\rm YM}
(1,2,3,4)\,,
\eea
therefore
\bea
E_1-E_2=(k_3\cdot k_4){\rm A}^\epsilon_{\rm YM}
(1,3,2,4)+(k_3\cdot k_4){\rm A}^\epsilon_{\rm YM}
(1,3,4,2)+(k_1\cdot k_2){\rm A}^\epsilon_{\rm YM}
(1,2,3,4)\,,~~~~\label{e1-e2}
\eea
where we employ $k_2\cdot (k_3+k_4)=-(k_2\cdot k_1)$.
Then, we apply the cyclic symmetry and ordered reserved identity again to get
\bea
\begin{split}
(k_3\cdot k_4){\rm A}^\epsilon_{\rm YM}
(1,3,2,4)+(k_3\cdot k_4){\rm A}^\epsilon_{\rm YM}
(1,3,4,2)~~~~\\
~~~=(k_3\cdot k_4){\rm A}^\epsilon_{\rm YM}
(4,1,3,2)+(k_3\cdot k_4){\rm A}^\epsilon_{\rm YM}
(4,3,1,2)\,~~~
\end{split}
\eea
To continue, we use the well known fundamental BCJ relation
\bea
(k_3\cdot (k_4+k_1)){\rm A}^\epsilon_{\rm YM}
(4,1,3,2)+(k_3\cdot k_4){\rm A}^\epsilon_{\rm YM}
(4,3,1,2)=0\,,
\eea
to arrive
\bea
(k_3\cdot k_4){\rm A}^\epsilon_{\rm YM}
(1,3,2,4)+(k_3\cdot k_4){\rm A}^\epsilon_{\rm YM}
(1,3,4,2)&=&-(k_3\cdot k_1){\rm A}^\epsilon_{\rm YM}
(4,1,3,2)\nn
&=&-(k_3\cdot k_1){\rm A}^\epsilon_{\rm YM}
(1,3,2,4)\,.
\eea
Then we use the fundamental BCJ relation
\bea
(k_3\cdot k_1){\rm A}^\epsilon_{\rm YM}
(1,3,2,4)+(k_3\cdot (k_1+k_2)){\rm A}^\epsilon_{\rm YM}
(1,2,3,4)=0\,
\eea
to obtain
\bea
(k_3\cdot k_4){\rm A}^\epsilon_{\rm YM}
(1,3,2,4)+(k_3\cdot k_4){\cal A}^\epsilon_{\rm YM}
(1,3,4,2)&=&(k_3\cdot (k_1+k_2)){\rm A}^\epsilon_{\rm YM}
(1,2,3,4)\nn
&=&-(k_3\cdot k_4){\rm A}^\epsilon_{\rm YM}
(1,2,3,4)\,.
\eea
Putting it back to \eref{e1-e2} we get
\bea
E_1-E_2=((k_1\cdot k_2)-(k_3\cdot k_4)){\rm A}^\epsilon_{\rm YM}
(1,2,3,4)\,.
\eea
Since
\bea
2k_1\cdot k_2=(k_1+k_2)^2=(k_3+k_4)^2=2k_3\cdot k_4\,,
\eea
we finally get
\bea
E_1-E_2=0\,,
\eea
thus although seems different, two expressions are indeed equivalent to each other.

Now we turn to the case photons carry flavors. Suppose there are two flavors labeled by ${{\bf 1}}$ and ${{\bf 2}}$ of external photons, ${{\bf 1}}$ is carried by $1$ and $3$, another one ${{\bf 2}}$ is carried by $2$ and $4$. Then, the corresponding partition is $\{(1_{{\bf 1}},3_{{\bf 1}}),(2_{{\bf 2}},4_{{\bf 2}})\}$, thus only the splitting $\{\{1_{{\bf 1}},3_{{\bf 2}},2_{{\bf 2}},4_{{\bf 1}}\}\}$ is allowed, which yields the
expansion
\bea
{\rm A}^{\epsilon,\W\epsilon}_{\rm EMf}(\{1_{{\bf 1}},2_{{\bf 2}},3_{{\bf 2}},4_{{\bf 1}}\};\emptyset)&=&-(k_2\cdot k_3){\rm A}^\epsilon_{\rm YM}
(1,3,2,4)\,.~~~~\label{result3}
\eea

Until now expansions in this subsection are given in the ordered splitting formula.
To get expansions into KK basis, one can identify coefficients of KK basis via the procedure
proposed in subsection \ref{coe-KK}. Since $1$ and $4$ are fixed at two ends in the color-ordering of KK basis,
there are two color-orderings $(1,2,3,4)$ and $(1,3,2,4)$ need to be considered. For the partitions $\{(1,2),(3,4)\}$, the ordered splitting
for color-ordering $(1,2,3,4)$ can be constructed as $\{\{1,2,3,4\}\}$, while the ordered splitting for ordering $(1,3,2,4)$ does not exist. For the
partition $\{(1,3),(2,4)\}$, the proper ordered splitting for color-ordering $(1,2,3,4)$ does not exist, while the splitting for color-ordering
$(1,3,2,4)$ can be found as $\{\{1,3,2,4\}\}$. For the partition $\{(1,4),(2,3)\}$, the ordered splitting for color-ordering $(1,2,3,4)$ does not exist,
while the splitting for color-ordering $(1,3,2,4)$ can be constructed as $\{\{1,4\},\{3,2\}\}$. Then coefficients for KK basis under each given partition
are
\bea
& &\{(1,2),(3,4)\}\,:~~~~C(2,3)=-(k_2\cdot k_3)\,,~~~~C(3,2)=0\,,\nn
& &\{(1,3),(2,4)\}\,:~~~~C(2,3)=0\,,~~~~C(3,2)=-(k_2\cdot k_3)\,,\nn
& &\{(1,4),(2,3)\}\,:~~~~C(2,3)=0\,,~~~~C(3,2)=-(k_3\cdot Z_3)\,,
\eea
where $C(2,3)$ and $C(3,2)$ denote the coefficients of ${\rm A}^\epsilon_{\rm YM}(1,2,3,4)$ and ${\rm A}^\epsilon_{\rm YM}(1,3,2,4)$.
Above results can be verified directly in formulae \eref{result1} and \eref{result3}.

\subsection{$5$-point EM amplitude ${\rm A}^{\epsilon,\W\epsilon}_{\rm EM}(\{1,5\};\{2,3,4\})$}

The second example is the $5$-point amplitude ${\rm A}^{\epsilon,\W\epsilon}_{\rm EM}(\{1,5\};\{2,3,4\})$
which includes two photons $1,5$ and three gravitons $2,3,4$. For this case, there is no need to distinguish whether photons carry flavor or not.
We chose the reference ordering as $5\prec1\prec3\prec4\prec2$.
Then we have following proper ordered splittings:
$\{\{1,5\},\{4,2,3\}\}$, $\{\{1,5\},\{2,4,3\}\}$,
$\{\{1,5\},\{4,3\},\{2\}\}$,$\{\{1,5\},\{2,3\},\{4\}\}$,
$\{\{1,5\},\{3\},\{2,4\}\}$, as well as $\{\{1,5\},\{3\},\{4\},\{2\}\}$. For these splittings, we have
\bea
& &G_0=1\,,~~H_1=\W\epsilon_3\cdot F_2\cdot F_4\cdot Z_4\,,\nn
& &G_0=1\,,~~H_1=\W\epsilon_3\cdot F_4\cdot F_2\cdot Z_2\,,\nn
& &G_0=1\,,~~H_1=\W\epsilon_3\cdot F_4\cdot Z_4,~~H_2=\W\epsilon_2\cdot Z_2\,,\nn
& &G_0=1\,,~~H_1=\W\epsilon_3\cdot F_2\cdot Z_2,~~H_2=\W\epsilon_4\cdot Z_4\,,\nn
& &G_0=1\,,~~H_1=\W\epsilon_3\cdot Z_3\,,~~H_2=\W\epsilon_4\cdot F_2\cdot Z_2\,,\nn
& &G_0=1\,,~~H_1=\W\epsilon_3\cdot Z_3\,,~~H_2=\W\epsilon_4\cdot Z_4\,,~~H_3=\W\epsilon_2\cdot Z_2\,,
\eea
and the expansion in the framework of ordered splitting expresses
\bea
{\rm A}^{\epsilon,\W\epsilon}_{\rm EM}(\{1,5\};\{2,3,4\})&=&(\W\epsilon_3\cdot f_2\cdot f_4\cdot Z_4){\rm A}^\epsilon_{\rm YM}
(1,4,2,3,5)+(\W\epsilon_3\cdot f_4\cdot f_2\cdot Z_2){\rm A}^\epsilon_{\rm YM}
(1,2,4,3,5)\nn
& &+\sum_{\shuffle}\,(\W\epsilon_3\cdot f_4\cdot Z_4)(\W\epsilon_2\cdot Z_2){\rm A}^\epsilon_{\rm YM}
(1,\{4,3\}\shuffle 2,5)\nn
& &+\sum_{\shuffle}\,(\W\epsilon_3\cdot f_2\cdot Z_2)(\W\epsilon_4\cdot Z_4){\rm A}^\epsilon_{\rm YM}
(1,\{2,3\}\shuffle 4,5)\nn
& &+\sum_{\shuffle}\,(\W\epsilon_4\cdot f_2\cdot Z_2)(\W\epsilon_3\cdot Z_3){\rm A}^\epsilon_{\rm YM}
(1,3\shuffle \{2,4\},5)\nn
& &+\sum_{\shuffle}\,(\W\epsilon_3\cdot Z_3)(\W\epsilon_4\cdot Z_4)(\W\epsilon_2\cdot Z_2){\rm A}^\epsilon_{\rm YM}
(1,3\shuffle 4\shuffle2,5)\,,
\eea
by using the general formula \eref{exp-EM3}.

Then we consider the expansion into KK basis. To illustrate the algorithm provided in subsection \ref{coe-KK},
we construct ordered splittings for the color-ordering $(1,2,4,3,5)$ in KK basis.
For the current case, there is only one partition $\{(1,5)\}$.
At the first step, the ordered subset ${\rm Or}^p_0$ which contain two photons $1$ and $5$ at two ends and two photons are nearby has only one candidate ${\rm Or}^p_0=\{1,5\}$.
Then, the lowest element in the set of remaining gravitons $\{1,2,3,4,5\}/{\rm Or}^p_0=\{2,3,4\}$ is $3$.
At the second step, we construct all possible subsets ${\rm Or}^p_1$ contain $3$ as its last element: ${\rm Or}^p_1=\{3\}$, ${\rm Or}^p_1=\{4,3\}$, ${\rm Or}^p_1=\{2,3\}$, ${\rm Or}^p_1=\{2,4,3\}$.  Notice that $\{4,2,3\}$ is not the correct choice of ${\rm Or}^p_1$
since it does not satisfy the color-ordering $(1,2,4,3,5)$ in KK basis. At the third step, we construct ${\rm Or}^p_2$ for all $\{1,2,3,4,5\}/{\rm Or}^p_0/{\rm Or}^p_1\neq\emptyset$. For ${\rm Or}^p_1=\{3\}$, the set of remaining elements is $\{1,2,3,4,5\}/{\rm Or}^p_0/{\rm Or}^p_1=\{2,4\}$, then ${\rm Or}^p_2=\{4\}$ and ${\rm Or}^p_2=\{2,4\}$ can be constructed. For ${\rm Or}^p_1=\{4,3\}$, we have ${\rm Or}^p_2=\{2\}$.
For ${\rm Or}^p_2=\{2,3\}$, we have ${\rm Or}^p_2=\{4\}$. At the final step, for ${\rm Or}^p_1=\{3\}$, ${\rm Or}^p_2=\{4\}$,
we have ${\rm Or}^p_3=\{2\}$. The above recursive construction can be summarized as
\bea
\{1,5\}\to\begin{cases} \displaystyle \{1,5\},\{3\}\to \begin{cases} \displaystyle \{1,5\},\{3\},\{4\}\to\{1,5\},\{3\},\{4\},\{2\}\\
\displaystyle \{1,5\},\{3\},\{2,4\}\end{cases}\\
\displaystyle \{1,5\},\{4,3\}\to \{1,5\},\{4,3\},\{2\}\\
\displaystyle \{1,5\},\{2,3\}\to \{1,5\},\{2,3\},\{4\}\\
\displaystyle \{1,5\},\{2,4,3\}\end{cases}
\eea
One can construct ordered splittings for other color-orderings in a similar way. After depicting $G_0$ and $H_l$
for each splitting, we get coefficients of KK basis as follows
\bea
C(2,3,4)&=&(\W\epsilon_3\cdot f_2\cdot Z_2)(\W\epsilon_4\cdot Z_4)\big|_{\{\{1,5\},\{2,3\},\{4\}\}}
+(\W\epsilon_4\cdot f_2\cdot Z_2)(\W\epsilon_3\cdot Z_3)\big|_{\{\{1,5\},\{3\},\{2,4\}\}}\nn
& &+(\W\epsilon_3\cdot Z_3)(\W\epsilon_4\cdot Z_4)(\W\epsilon_2\cdot Z_2)\big|_{\{\{1,5\},\{3\},\{4\},\{2\}\}}\,,\nn
C(2,4,3)&=&\W\epsilon_3\cdot f_4\cdot f_2\cdot Z_2\big|_{\{\{1,5\},\{2,4,3\}\}}+(\W\epsilon_3\cdot f_4\cdot Z_4)(\W\epsilon_2\cdot Z_2)\big|_{\{\{1,5\},\{4,3\},\{2\}\}}\nn
& &+(\W\epsilon_3\cdot f_2\cdot Z_2)(\W\epsilon_4\cdot Z_4)\big|_{\{\{1,5\},\{2,3\},\{4\}\}}+(\W\epsilon_4\cdot f_2\cdot Z_2)(\W\epsilon_3\cdot Z_3)\big|_{\{\{1,5\},\{3\},\{2,4\}\}}\nn
& &+(\W\epsilon_3\cdot Z_3)(\W\epsilon_4\cdot Z_4)(\W\epsilon_2\cdot Z_2)\big|_{\{\{1,5\},\{3\},\{4\},\{2\}\}}\,,\nn
C(3,2,4)&=&(\W\epsilon_4\cdot f_2\cdot Z_2)(\W\epsilon_3\cdot Z_3)\big|_{\{\{1,5\},\{3\},\{2,4\}\}}+(\W\epsilon_3\cdot Z_3)(\W\epsilon_4\cdot Z_4)(\W\epsilon_2\cdot Z_2)\big|_{\{\{1,5\},\{3\},\{4\},\{2\}\}}\,,\nn
C(3,4,2)&=&(\W\epsilon_3\cdot Z_3)(\W\epsilon_4\cdot Z_4)(\W\epsilon_2\cdot Z_2)\big|_{\{\{1,5\},\{3\},\{4\},\{2\}\}}\,,\nn
C(4,2,3)&=&\W\epsilon_3\cdot f_2\cdot f_4\cdot Z_4\big|_{\{\{1,5\},\{4,2,3\}\}}+(\W\epsilon_3\cdot f_4\cdot Z_4)(\W\epsilon_2\cdot Z_2)\big|_{\{\{1,5\},\{4,3\},\{2\}\}}\nn
& &+(\W\epsilon_3\cdot f_2\cdot Z_2)(\W\epsilon_4\cdot Z_4)\big|_{\{\{1,5\},\{2,3\},\{4\}\}}+(\W\epsilon_3\cdot Z_3)(\W\epsilon_4\cdot Z_4)(\W\epsilon_2\cdot Z_2)\big|_{\{\{1,5\},\{3\},\{4\},\{2\}\}}\,,\nn
C(4,3,2)&=&(\W\epsilon_3\cdot f_4\cdot Z_4)(\W\epsilon_2\cdot Z_2)\big|_{\{\{1,5\},\{4,3\},\{2\}\}}+(\W\epsilon_3\cdot Z_3)(\W\epsilon_4\cdot Z_4)(\W\epsilon_2\cdot Z_2)\big|_{\{\{1,5\},\{3\},\{4\},\{2\}\}}\,,~~~~~~
\eea
where $C(i,j,k)$ denotes the coefficient of the color-ordered YM amplitude ${\rm A}^\epsilon_{\rm YM}(1,i,j,k,5)$. Since the combinatory momentum $Z_i$ depends not only on the color-ordering but also on the ordered splitting , we have explicitly wrote down the ordered splitting for each term.

\subsection{$5$-point EM amplitude ${\rm A}^{\epsilon,\W\epsilon}_{\rm EM}(\{1,3,4,5\};\{2\})$}

The final example is the $5$-point amplitude ${\rm A}^{\epsilon,\W\epsilon}_{\rm EM}(\{1,3,4,5\};\{2\})$
which contains four photons $1,3,4,5$ and one graviton $2$. For this amplitude, we still chose the reference ordering $5\prec1\prec3\prec4\prec2$.
We start with the case photons do not carry flavor. Then the allowed ordered splittings for each partition can be found as follows:
For the partition $\{(1,5),(3,4)\}$, the proper splittings are $\{\{1,5\},\{2,4,3\}\}$ and
$\{\{1,5\},\{4,3\},\{2\}\}$; For the partition $\{(1,4),(3,5)\}$, the proper splittings are $\{\{1,4,2,3,5\}\}$ and
$\{\{1,4,3,5\},\{2\}\}$; For the partition $\{(1,3),(4,5)\}$, the proper splittings are $\{\{1,3,2,4,5\}\}$ and $\{\{1,3,4,5\},\{2\}\}$.
For these ordered splittings, we have
\bea
& &G_0=1,~~H_1=k_4\cdot f_2\cdot Z_2\,,\nn
& &G_0=1,~~H_1=k_4\cdot Z_4,~~H_2=\W\epsilon_2\cdot Z_2\,,\nn
& &G_0=k_4\cdot f_2\cdot k_3\,,\nn
& &G_0=k_4\cdot k_3,~~H_1=\W\epsilon_2\cdot Z_2\,,\nn
& &G_0=k_3\cdot f_2\cdot k_4\,,\nn
& &G_0=k_3\cdot k_4,~~H_1=\W\epsilon_2\cdot Z_2\,.
\eea
Thus, applying the general formula \eref{exp-EM3}, we get the expansion in the ordered splitting formula
\bea
{\rm A}^{\epsilon,\W\epsilon}_{\rm EM}(\{1,3,4,5\};\{2\})&=&-(k_4\cdot f_2\cdot Z_2){\rm A}^\epsilon_{\rm YM}
(1,2,4,3,5)-\sum_{\shuffle}\,(k_4\cdot Z_4)(\W\epsilon_2\cdot Z_2){\rm A}^\epsilon_{\rm YM}
(1,\{4,3\}\shuffle 2,5)\nn
& &-(k_4\cdot f_2\cdot k_3){\rm A}^\epsilon_{\rm YM}
(1,4,2,3,5)-\sum_{\shuffle}\,(k_4\cdot k_3)(\W\epsilon_2\cdot Z_2){\rm A}^\epsilon_{\rm YM}
(1,\{4,3\}\shuffle 2,5)\nn
& &-(k_3\cdot f_2\cdot k_4){\rm A}^\epsilon_{\rm YM}
(1,3,2,4,5)-\sum_{\shuffle}\,(k_3\cdot k_4)(\W\epsilon_2\cdot Z_2){\rm A}^\epsilon_{\rm YM}
(1,\{3,4\}\shuffle 2,5)\,.~~~~~
\eea

Then we turn to the case photons carry flavors. Assuming there are two flavors labeled by ${\bf 1}$ and ${\bf 2}$, ${\bf 1}$ is carried by photons $1$
and $5$, while ${\bf 2}$ is carried by $3$ and $4$. One can find that the proper ordered splittings for the current case has only two candidates
$\{\{1_{{\bf 1}},5_{{\bf 1}}\},\{2,4_{{\bf 2}},3_{{\bf 2}}\}\}$ and $\{\{1_{{\bf 1}},5_{{\bf 1}}\},\{4_{{\bf 2}},3_{{\bf 2}}\},\{2\}\}$. Then the expansion in the ordered splitting formula is given as
\bea
{\rm A}^{\epsilon,\W\epsilon}_{\rm EMf}(\{1_{{\bf 1}},3_{{\bf 2}},4_{{\bf 2}},5_{{\bf 1}}\};\{2\})&=&-(k_4\cdot F_2\cdot Z_2){\rm A}^\epsilon_{\rm YM}
(1,2,4,3,5)\nn & &-\sum_{\shuffle}\,(k_4\cdot Z_4)(\W\epsilon_2\cdot Z_2){\rm A}^\epsilon_{\rm YM}
(1,\{4,3\}\shuffle 2,5)\,.
\eea

Finally, we consider the coefficients of KK basis. We choose the color-ordering $(1,2,4,3,5)$ as the example to illustrate the algorithm
proposed in subsection \ref{coe-KK}. The recursive construction of ordered splittings can be summarized as follows,
\begin{itemize}
\item Partition $\{(1,5),(3,4)\}$:
\bea
\{1,5\}\to\begin{cases} \displaystyle \{1,5\},\{4,3\}\to \{1,5\},\{4,3\},\{2\}\\
\displaystyle \{1,5\},\{2,4,3\}\end{cases}
\eea
At the first step, $\{1,5\}$ is the only choice that ${\rm Or}^p_0$ contain photons $1$ and $5$ at two ends and two photons are nearby.
At the second step, the choices $\{1,5\},\{2,3\}$ and $\{1,5\},\{4,2,3\}$ are excluded since the photons $3$ and $4$ should appear at nearby positions in
the same subset, due to the pair $(3,4)$ in the partition. The second one $\{1,5\},\{4,2,3\}$ also violates the desired color-ordering $(1,2,4,3,5)$.
\item Partition $\{(1,4),(3,5)\}$:
\bea
\{1,4,3,5\}\to\{1,4,3,5\},\{2\}
\eea
At the first step, we drop the candidate ${\rm Or}^p_0=\{1,4,2,3,5\}$ which violates the color-ordering $(1,2,4,3,5)$. 
\item Partition $\{(1,3),(4,5)\}$:
The correct ordered splitting does not exist since
both ${\rm Or}^p_0=\{1,3,4,5\}$ and ${\rm Or}^p_0=\{1,3,2,4,5\}$ voilate the color-ordering $(1,2,4,3,5)$.
\end{itemize}

Before ending this subsection, we list the coefficients of KK basis for each partition:
\begin{itemize}
\item Partition $\{(1,5),(3,4)\}$:
\bea
C(2,3,4)&=&0\,,~~~~
C(3,2,4)=0\,,~~~~C(3,4,2)=0\,,\nn C(4,2,3)&=&-(k_4\cdot Z_4)(\W\epsilon_2\cdot Z_2)\big|_{\{\{1,5\},\{4,3\},\{2\}\}}\,,~~~~C(4,3,2)=-(k_4\cdot Z_4)(\W\epsilon_2\cdot Z_2)\big|_{\{\{1,5\},\{4,3\},\{2\}\}}\,,\nn
C(2,4,3)&=&-k_4\cdot f_2\cdot Z_2\big|_{\{\{1,5\},\{2,4,3\}\}}-(k_4\cdot Z_4)(\W\epsilon_2\cdot Z_2)\big|_{\{\{1,5\},\{4,3\},\{2\}\}}\,.
\eea
\item Partition $\{(1,4),(3,5)\}$:
\bea
C(2,3,4)&=&0\,,~~~~C(3,2,4)=0\,,~~~~C(3,4,2)=0\,,\nn
C(2,4,3)&=&-(k_4\cdot k_3)(\W\epsilon_2\cdot Z_2)\big|_{\{\{1,4,3,5\},\{2\}\}}\,,~~~~C(4,3,2)=-(k_4\cdot k_3)(\W\epsilon_2\cdot Z_2)\big|_{\{\{1,4,3,5\},\{2\}}\,,\nn
C(4,2,3)&=&-k_4\cdot f_2\cdot k_3\big|_{\{\{1,4,2,3,5\}\}}-(k_4\cdot k_3)(\W\epsilon_2\cdot Z_2)\big|_{\{\{1,4,3,5\},\{2\}}\,.
\eea
\item Partition $\{(1,3),(4,5)\}$:
\bea
C(2,4,3)&=&0\,,~~~~C(4,2,3)=0\,,~~~~C(4,3,2)=0\,,\nn
C(2,3,4)&=&-(k_3\cdot k_4)(\W\epsilon_2\cdot Z_2)\big|_{\{\{1,3,4,5\},\{2\}\}}\,,~~~~C(3,4,2)=-(k_3\cdot k_4)(\W\epsilon_2\cdot Z_2)\big|_{\{\{1,3,4,5\},\{2\}\}}\,,\nn
C(3,2,4)&=&-k_3\cdot f_2\cdot k_4\big|_{\{\{1,3,2,4,5\}\}}-(k_3\cdot k_4)(\W\epsilon_2\cdot Z_2)\big|_{\{\{1,3,4,5\},\{2\}\}}\,.
\eea
\end{itemize}
%

\section{Summary}
\label{conclusion}

We demonstrate how to obtain the expansion of tree EM amplitudes into KK basis of tree YM amplitudes efficiently by applying proper differential operators in this paper. The coefficients for KK basis in the expansion are shared by the expansion of tree $\phi^4$ amplitudes into tree BS amplitudes, the expansion of tree sYMS amplitudes into tree BS amplitudes, as well the expansion of tree DBI amplitudes into tree special extended DBI amplitudes, as have been explained in detail.
These expansions exhibit the connections among amplitudes of different theories which are invisible from the angle of Feynman rules,
and serve as the dual representations of unifying relations described by differential operators.

The method used in \cite{Feng:2019tvb} and this paper can also be applied to other theories linked by differential operators.
One of our future direction is to derive expansions of other theories via this method and constructe a complete web for expansions.

Interestingly, for expansions of EM amplitudes obtained in the current paper, the manifest gauge invariance
is missing for all gravitons\footnote{When saying "gauge invariance" for a particular particle $i$, we means the Ward's identity that the amplitude
vanishes under the replacement $\W\epsilon_i\to k_i$.}. The lose of manifest gauge invariance is a general feature for expansions of amplitudes
into KK basis. As discussed in \cite{Feng:2019tvb}, for the expansion of single-trace EYM amplitudes, the manifest gauge invariance
for all gravitons can be ensured when expanding into BCJ basis rather than KK basis, with the cost that coefficients contain poles.
For EM amplitudes, how to reproduce the manifest gauge invariance for gravitons is a significative problem.

The expansions of amplitudes not only provide the theoretical understanding of connections between different theories, but also
benefits the practical calculations. For example, since the evaluation of YM amplitudes is much easier than that of EM amplitudes,
one can calculate YM amplitudes at the first step, and get EM amplitudes through the expansions. The obtained EM amplitudes
may be used to study the quantum corrections of the behavior of photons in the gravitational field, such as gravitational light bending
and Hawking radiation.

\section*{Acknowledgments}

We would like to thank Prof. Bo Feng and Xiaodi Li for valuable discussions.
Shi-Qian Hu is supported by the Postgraduate Reaearch \& Practice Innovation Program of Jiangsu Province (KYCX19\_2098).
Kang Zhou is supported by Chinese NSF funding under
No.11805163, as well as NSF of Jiangsu Province under Grant No.BK20180897.

\end{document}